\documentclass[sigconf]{acmart}

\usepackage{subcaption}

\copyrightyear{2025}
\acmYear{2025}
\setcopyright{cc}
\setcctype{by}
\acmConference[Websci '25]{Proceedings of the 17th ACM Web Science Conference 2025}{May 20--24, 2025}{New Brunswick, NJ, USA}
\acmBooktitle{Proceedings of the 17th ACM Web Science Conference 2025 (Websci '25), May 20--24, 2025, New Brunswick, NJ, USA}
\acmDOI{10.1145/3717867.3717886}
\acmISBN{979-8-4007-1483-2/2025/05}

\usepackage{xcolor}

\newcommand{\cljb}[1]{#1}

\begin{document}

\title{A Call to Arms: Automated Methods for Identifying Weapons in Social Media Analysis of Conflict Zones}

\author{Afia Fairoose Abedin}
\affiliation{%
 \institution{University of Alberta}
  \country{Canada}
 }
 
 \author{Abdul Bais}
\affiliation{%
 \institution{University of Regina}
  \country{Canada}
 }

 \author{Cody Buntain}
\affiliation{%
 \institution{University of Maryland}
  \country{USA}
 }

 \author{Laura Courchesne}
\affiliation{%
 \institution{Stanford University}
  \country{USA}
 }

 \author{Brian McQuinn}
\affiliation{%
 \institution{University of Regina}
  \country{Canada}
 }

 \author{Matthew E. Taylor}
\affiliation{%
 \institution{University of Alberta}
  \country{Canada}
 }

 \author{Muhib Ullah}
\affiliation{%
 \institution{University of Regina}
  \country{Canada}
 }

\begin{abstract}
The massive proliferation of social media data represents a transformative opportunity for conflict studies and for tracking the proliferation and use of weaponry, as conflicts are increasingly documented in these online spaces. 
At the same time, the scale and types of data available are problematic for traditional open-source intelligence.
This paper focuses on identifying specific weapon systems and the insignias of the armed groups using them as documented in the Ukraine war, as these tasks are critical to operational intelligence and tracking weapon proliferation, especially given the scale of international military aid given to Ukraine. 
The large scale of social media makes manual assessment difficult, however, so this paper presents early work that uses computer vision models to support this task.
We demonstrate that these models can both identify weapons embedded in images shared in social media and how the resulting collection of military-relevant images and their post times interact with the offline, real-world conflict.
Not only can we then track changes in the prevalence of images of tanks, land mines, military trucks, etc., we find correlations among time series data associated with these images and the daily fatalities in this conflict.
This work shows substantial opportunity for examining similar online documentation of conflict contexts, and we also point to future avenues where computer vision can be further improved for these open-source intelligence tasks.
\end{abstract}



\maketitle

\begin{figure*}[t]
\begin{center}
\includegraphics[width=0.95\textwidth]{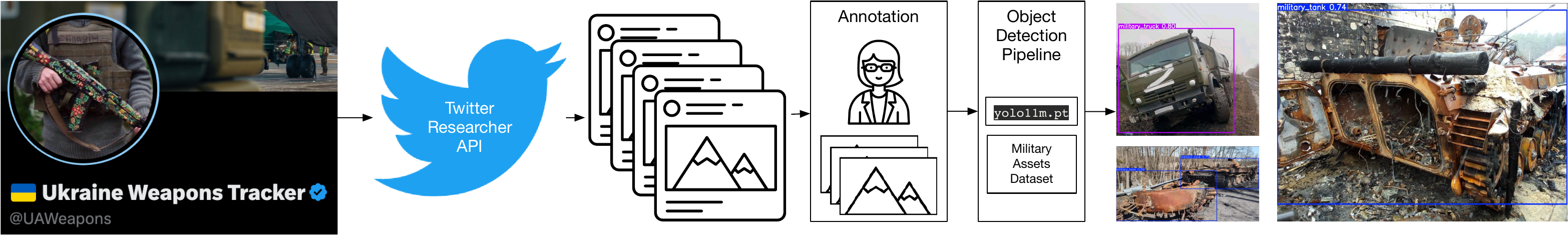}
\caption{Pipeline for Collecting, Annotating, and Automatically Identifying Weapons in Social Media Content}
\label{fig:pipeline}
\end{center}
\end{figure*}

\section{Introduction}

The Russian military invaded Ukraine on 24 February 2022. 
The war has devastated the country's infrastructure, killed hundreds of thousands of people, and displaced more than 6.7 million people according to the United Nations High Commissioner for Refugees.\footnote{\url{https://data.unhcr.org/en/situations/ukraine}} 
Despite Russia's predictions of a quick victory, the war is entering its fourth year.
Ukraine's military heavily depends on weapons and munitions from North Atlantic Treaty Organization (NATO) countries, with billions in humanitarian and military support from the US, Canada, and other NATO countries. 
A vital component of this aid includes advanced weapon systems, such as the Javelin shoulder-fired anti-tank system, long-range mobile missile launchers, and controversial cluster munitions. 
While Ukraine's forces make heavy use of this materiel, their patchwork military consists of units and non-state militias (with varying degrees of independence from the state). 
This irregular framework complicates our understanding of the flow and movement of weapons.
These complications, coupled with the minimal effort put into tracking these weapons, have left NATO officials in the fog of war, with little understanding of these systems' deployment and use. 
This article provides a novel path through this fog via social media analysis, specifically an investigation of the @UAWeapons Twitter/X account, which compiled  images and video from various sources documenting the ``usage/capture of materiel in Ukraine'' between February 2022 and October 2023.\footnote{https://x.com/uaweapons} The analysis used only this account to test the proposed methods. This choice limits the results but future analysis will expand the dataset.

Examining this open-source intelligence covering the war in Ukraine presents new opportunities for documentation in digital spaces.
Historically, operational security tenets have banned soldiers from posting on social networks in war zones, as opponents can use this media for intelligence. 
But in the Ukraine war, Ukrainian President Volodymyr Zelensky has driven a sustained effort to use social media to solicit international support for Ukraine’s cause and counterbalance substantial Russian disinformation campaigns. 
The result is an expansive and real-time record of the events of the war captured through videos, photographs, and reports uploaded to YouTube, Facebook, Instagram, Telegram, and Twitter/X. Following the large-scale use of social media in this conflict, the scale of the data being produced is similarly unprecedented, as the uploading of videos and photographs makes this the most documented war in history \citep{videoNPR}. 

This paper presents a multidisciplinary descriptive study that combines computer vision and armed group analysis to identify new insights and connections between kinetic warfare in conflict zones and digital-trace data. 
By training computer vision systems to identify weapon systems and insignias of armed groups in the @UAWeapons coverage, we can measure the frequency of such use and compare its prevalence in online coverage to actual kinetic engagements and casualties of the conflict.
We demonstrate, through analyzing 5,601 social media posts and 7,060 images posted between 21 February 2022 and 31 December 2022, that audiences respond differentially across various types of weapons identified via YOLOv11 \citep{YOLO16}.
Further building on this analysis of sharing images of weapons, we compare time series data of daily prevalence of weapon types to armed clashes and casualty data from the Armed Conflict Location and Event Data (ACLED) \citep{Raleigh:2023aa}.
While overall frequency of UAWeapons posts show no significant connection with this offline conflict data, we do find significant relationships between daily casualty rates and images showing land mines, civilian vehicles, and military trucks.
Specifically, images of land mines appear to lead fatalities by four days, and fatalities lead images of military trucks and camouflaged soldiers by four and six days respectively.
Connecting these findings back to audience engagement, these image-types do not receive the most engagement, suggesting their prevalence online is not driven by audience interest but potentially by offline events.

These results demonstrate the utility of digital-trace data in penetrating the fog of war and providing additional insight into conflicts.
As concerns around the war in Ukraine heighten, along with increased concern for conflict in Taiwan and climate-related conflict, harvesting such new insights will likewise be of increasing value for maintaining situational awareness.

\section{Related Work}


Social media has been a key component of open-source intelligence for over a decade, from early work in crisis informatics \citep{10.1145/1240624.1240736.2007,2011}---where emergency responders use social media data to enhance situational awareness during disasters---to studies of public response to terrorism \citep{Buntain_Golbeck_Liu_LaFree_2021} to the spread of COVID-19 \citep{info:doi/10.2196/19273}.
Much of this work has demonstrated the prevalence of high-priority and useful information shared online during crises \citep{2022}, with recent work demonstrating the unique insights social media provides into conflict, as in the Syrian civil war \citep{10.1177/1750635216653903.2017} or journalists' use of social media in Iraq, Yemen, and elsewhere \citep{10.1080/17512786.2021.1908839.2023}.
To this end, Gambo \cite{10.1007/978-3-030-62183-4_33.2021} argues that the proliferation of small arms in Egypt is an artifact of the social media-driven Arab Spring,  while Picard \cite{10.1007/978-3-030-65636-2_8.2021} introduces social media explicitly as a source of relevant data for tracking small arms, a process on which Harvey et al.~\cite{Harvey_Lebret_Massonnet_Aberer_Demartini_2023} improves for tracking firearms using Twitter data.
Together, these works demonstrate social media's potential utility as a means to understand the flow of weaponry, small arms, and materiel in conflict settings.

Although government tracking of weapons is generally not made public, there is a long history of non-government organizations tracking global weapons trade with a particular focus on small arms tracking in conflict zones \citep{smallArms}.
We partnered with the Small Arms Survey in Geneva, Switzerland  to train computer vision algorithms to identify four weapon systems or sub-categories of weapons. 
As a starting point, the team selected: javelin antitank missiles (FGM-148 AAWS-M), landmines, grenades, and rocket-propelled grenades (RPGs). 
These systems are difficult to identify in real-world pictures, and landmines are of particular humanitarian concern. Additionally, the javelin antitank missile has impacted the course of the war in significant ways~\citep{Javelin}. The team also trained the system to identify insignias worn by combatants, which can help determine which military or non-state armed groups possess the weapons.

This research builds on our previous collaborations, bringing together experts in crisis informatics, armed violence, and AI. Previously, the team conducted the most comprehensive study of social media in Afghanistan  [\emph{blind review}], highlighting how technology can offer a real-time record of conflict events (e.g., troop movements, key battles, and levels of community support). 


This paper focuses on the Ukraine conflict. There has been an increase in paramilitary groups involved in the conflict on both sides since the annexation of Crimea by Russia in February 2014. The full-scale invasion of Russia in February 2022 has only amplified this dynamic. Paramilitary groups are defined as military combat groups that are aligned with the Ukrainian or Russian government, but not directly incorporated into the military chain of command. 

A computer vision system using machine learning and advanced algorithms can detect and measure weapons in social media images at scale more effectively than manual approaches. Object detection and classification are essential for weapon tracking, and deep learning (DL) has shown considerable progress in these areas. DL-based techniques, specifically convolutional neural networks (CNNs), have dominated the field of smart surveillance and security systems \citep{nikouei2018,xu2018,narejo2021,iqbal2021}. Hashmi et al.~\cite{hashmi2021} studied automating weapon detection using CNN-based DL models. They compared You Only Look Once~\citep{YOLO16} (YOLO) versions 3 and 4, highlighting the advancements in processing speed and sensitivity achieved by YOLOv4. Bajon \cite{bajon2022} used DL for small arms recognition, and Hnoohom et al.~\cite{hnoohom2022} used image-tiling-based DL to detect small weapons.

\section{Methods}

As outlined in Figure \ref{fig:pipeline}, the analyses herein rely on open-source intelligence collected from social media, Twitter in particular, and uses a fine-tuned object-detection model to identify military-related objects at scale.
Using output from these models, we then assess whether specific types of weapon systems inform us about events, armed clashes, and fatalities on the ground across 2022 and Russia's main invasion of Ukraine.

\subsection{Datasets}

This analysis leverages three datasets: 1) a collection of Twitter posts from the @UAWeapons account, 2) a manually labeled collection of images annotated with five types of military-relevant objects, and 3) an extant dataset of images depicting twelve types of military assets \cite{Madhuwala:2024aa}.

\begin{figure}[ht]
\begin{center}
\includegraphics[width=0.45\textwidth]{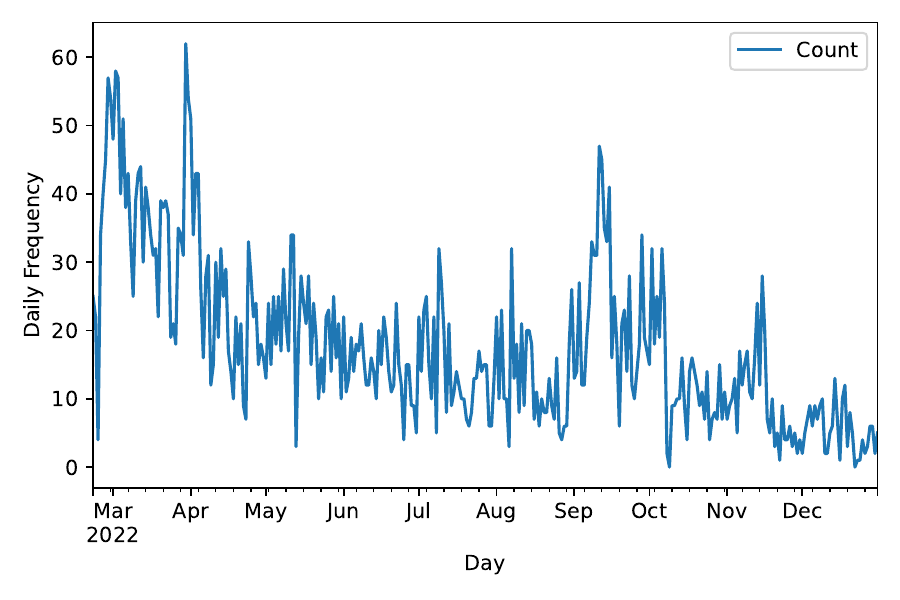}
\caption{Daily Post Frequency for @UAWeapons. Sum total of 5,601 posts in 2022.}
\label{fig:post_freq}
\end{center}
\end{figure}

\paragraph{Social Media Data}
Military-related blogs---``milblogs'' for short \citep{Resteigne01082010}---have become popular during the war in Ukraine.
While some analysis has suggested much of the milblog space is Russian-sponsored, milbloggers exist on both sides of the conflict, providing increasingly technical coverage of the conflict \cite{10.1007/s42001-023-00240-9}.
On Twitter, one such account, @UAWeapons, launched in February 2022 at the beginning of the Russian invasion of Ukraine, has provided substantial coverage of military materiel in the country, including a substantial volume of images. The account compiles images from several accounts to centralize this type of data. We selected this account because it was a compilation, and it provided a sufficiently large dataset to test our methods and hypotheses.
Using Twitter's now-defunct researcher API, we have collected all 5,601 tweets posted by this account between its inception on 21 February 2022 and the end of that year (Figure \ref{fig:post_freq}).
We performed this retrospective collection on 20 January 2023.
From these social media posts, we then collected all images the account shared during this timeframe, resulting in 7,060 images.

\paragraph{Manual Image Annotation} 
While the tweets contain text around our collected images, these texts do not often describe the specific object in the image. 
To support the automatic identification of military-related images, we therefore manually labeled a sample of these images, outlining bounding boxes in each image and whether that box contains one of the following objects:

\begin{itemize}
\item A grenade,
\item A land mine,
\item A rocket-propelled grendate (RPG),
\item An anti-tank weapon (e.g., the US-provided Javelin shoulder-mounted anti-tank weapon), or
\item A military/militia insignia.
\end{itemize}

To this end, we sampled 567 images from our collection and selected and labeled these images, focusing on samples that capture diverse scenarios and weapon classes. 
As these images are sourced from social media, they present several quality challenges, such as low resolution, dynamic environments, blurring, occlusion, varying distance from the camera, and reduced visibility of weapons. 
Table \ref{tab:object_freq} shows the frequencies of each object; one should note that an image may contain multiple objects, so the number of objects annotated does not equal the number of images.

\begin{table}[htp]
\caption{Frequency of Manually Annotated Military Objects}
\begin{center}
\begin{tabular}{l r}

\textbf{Object} & \textbf{Frequency} \\ \hline
RPG & 279 \\
Insignia & 128 \\
Grenade & 95 \\
Land Mine & 44 \\
Anti-Tank & 43 \\

\end{tabular}
\end{center}
\label{tab:object_freq}
\end{table}%

\begin{table}[htb]
\caption{Military Objects in the Supplemental Dataset}
\begin{center}
\begin{tabular}{l r}

\textbf{Object} &  \textbf{Frequency} \\ \hline

Military Tank   &   20,059   \\
Military Aircraft   &   8,741    \\
Soldier &   7,807    \\
Camouflage Soldier  &   5,499    \\
Military Vehicle    &   2,464    \\
Military Warship    &   2,425    \\
Weapon  &   1,568    \\
Military Truck  &   1,489    \\
Military Artillery  &   606 \\
Civilian Vehicle    &   586 \\
Civilian    &   53  \\
Trench  &   44 

\end{tabular}
\end{center}
\label{tab:kaggle_freq}
\end{table}%

\paragraph{Supplemental Images of Military Assets} 
Our manually annotated collection of images is relatively small for fine-tuning extant object-detection models.
As such, we also rely on a supplementary dataset containing 26,315 images of military assets, specifically one containing twelve types of images \citep{Madhuwala:2024aa}---see Table \ref{tab:kaggle_freq} for a breakdown of image types.
We use these images to expand our manually annotated collection, resulting in a final dataset of 27,386 images.


\begin{figure*}[hbtp]
\centering
    \begin{subfigure}[t]{0.5\textwidth}
        \centering
        \includegraphics[width=0.95\textwidth]{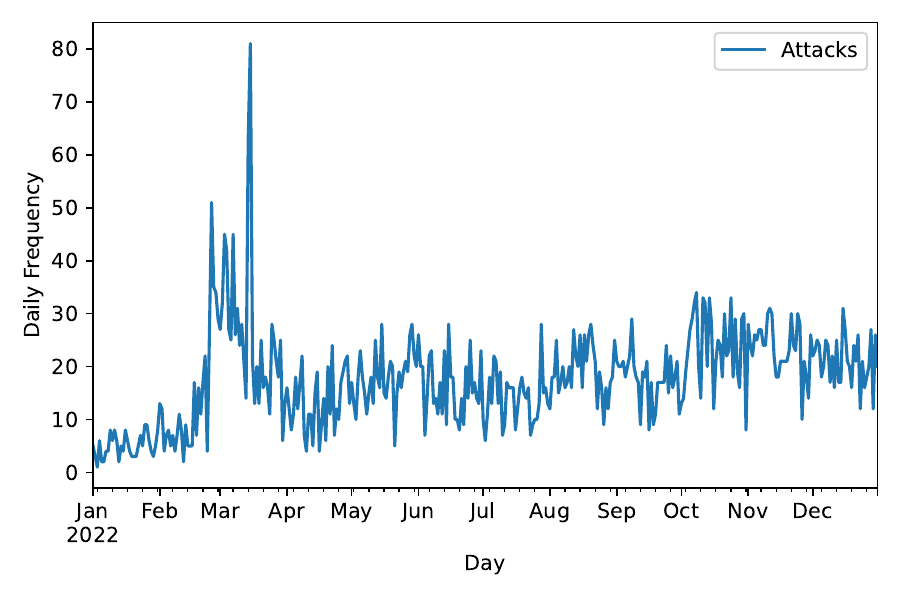}
        \caption{Attacks in Ukraine Per Day}\label{fig:acled_freqs_attack}
    \end{subfigure}%
    ~ 
    \begin{subfigure}[t]{0.5\textwidth}
        \centering
        \includegraphics[width=0.95\textwidth]{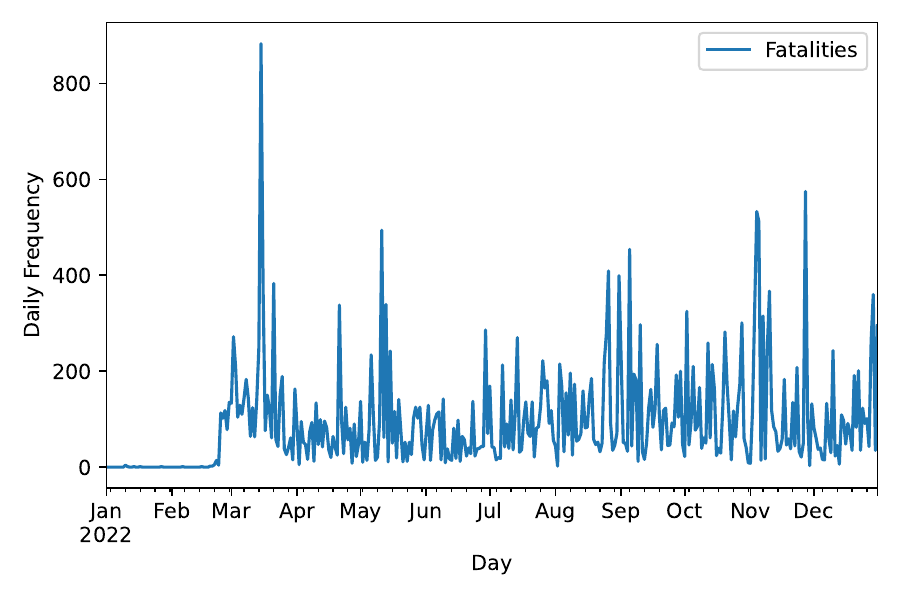}
        \caption{Fatalities in Ukraine Per Day}\label{fig:acled_freqs_fatal}
    \end{subfigure}
    \caption{ACLED Conflict Data for Ukraine \& the Black Sea.}
\label{fig:acled_freqs}
\end{figure*}

\subsection{Methods for Object Detection}

YOLO (You Only Look Once)~\citep{YOLO16} is an open-source, real-time object detection network commonly used in computer vision tasks. 
Its architecture consists of a backbone, a neck, and a head. 
The backbone determines feature representation and affects inference efficiency. 
The neck combines physical and semantic features to create feature maps. The head predicts detection outcomes using these multi-level features. 
The backbone's design greatly affects the model's effectiveness and efficiency, while the neck's integration of features from multiple scales is crucial for object detection. 

YOLOv11, the latest version in the YOLO series introduced in mid-2024, incorporates innovative modifications such as self-attention, feature fusion, and context aggregation elements to improve speed and accuracy in object detection \citep{jocher2023}. 
The model's proficiency in detecting objects of varying scales is due to a feature pyramid network~\cite{lin2017}. 
These improvements in the architecture of YOLOv11 makes it suitable for detecting challenging images, especially when the objects are in occlusion states, which is often the case in different weapons. 
Building on the YOLOv11 pre-trained model, we then fine-tune this model using the annotated and extended datasets described above, learning for 200 epochs.

\subsection{Correlating Content to Offline Conflict}

While the above methods provide insights into the online space and prevalence of various pieces of military materiel, to gain insight into the on-the-ground conflict, we rely on the ACLED dataset \citep{Raleigh:2023aa}.
ACLED provides fine-grained data on daily events within the Ukraine conflict as part of its ``Ukraine \& the Black Sea'' conflict dataset.
From this dataset, we extract two measures: 1) the number of armed clashes/attacks per day (Figure \ref{fig:acled_freqs_attack}), and 2) the number of fatalities per day (Figure \ref{fig:acled_freqs_fatal}).

To assess connections between the online behavior and offline event data, we first transform our online data into two sets: raw daily counts of the various image types, and a normalized transformation such that we measure the percentage of each image type in a given day.
We then compare these two time series with daily attack and fatality data using first Granger causality and then vector autoregression (VAR).

In both methods, we measure the degree to which one time series predicts another and assess the temporal relationships between these two time series via lags.
For example, we assess whether the number of tank-images per day predicts a lagged version of the number of fatalities per day, under the expectation that a higher frequency of tanks on day 1 may indicate an upcoming attack on day 2 or day 3.
For both Granger causality and VAR, we use a maximum lag of seven days, and we assess relationships in online-to-offline and offline-to-online directions.
While these relationships are not strictly causal in the purest sense, Granger causality and VAR methods have been used previously to understand such relationships using online data, as in \citet{10.1080/10584609.2019.1661889}.

\section{Results}

\subsection{Military-Relevant Object Detection}

Mean average precision (mAP) is a common metric in object detection tasks, assessing the performance of models across various images of objects by considering both precision (P) and recall (R). 
P represents how accurate the model's predictions are for a particular class, while R measures how well the model captures all relevant objects of that class. 
To calculate mAP, we compute the average precision (AP) for each individual object class, considering true positives, false positives, and false negatives. 
These AP scores are then averaged across all classes to yield the mAP.

\begin{table}[htb]
\caption{Detection Performance Across Military Objects.}
\begin{center}
\begin{tabular}{l r r r r r}

\textbf{Class}  &  \textbf{Instances}  &  \textbf{Precision}  &  \textbf{Recall}  &  \textbf{mAP} \\ \hline 
Military Tank  & 1,787 & 0.797 & 0.905 &  0.892 \\ 
Military Aircraft  & 1,063 & 0.905 & 0.865 &  0.915 \\ 
Camouflage Soldier  & 633 & 0.809 & 0.65 &  0.698 \\ 
Soldier  & 745 & 0.805 & 0.683 &  0.771 \\ 
Weapon  & 358 & 0.782 & 0.617 &  0.673 \\ 
Military Vehicle  & 307 & 0.604 & 0.485 &  0.599 \\ 
Military Warship  & 177 & 0.508 & 0.887 &  0.513 \\ 
Grenade  & 237 & 0.879 & 0.751 &  0.863 \\ 
Military Artillery  & 117 & 0.726 & 0.513 &  0.659 \\ 
Military Truck  & 148 & 0.571 & 0.689 &  0.664 \\ 
Insignia  & 31 & 0.240 & 0.097 &  0.094 \\ 
Civilian Vehicle  & 42 & 0.576 & 0.389 &  0.465 \\ 
RPG  & 17 & 0.601 & 1 &  0.924 \\ 
Civilian  & 1 & 0 & 0 &  0 \\ 
Trench  & 3 & 0 & 0 &  0 \\ 
\hline 
All  & 5,666 & 0.587 & 0.569 &  0.582 \\ 

\end{tabular}
\end{center}
\label{tab:yolo_perf}
\end{table}

\begin{figure*}[hbtp]
\centering
\begin{tabular}{c@{\quad}c@{\quad}c@{\quad}c}
\includegraphics[width=.22\textwidth]{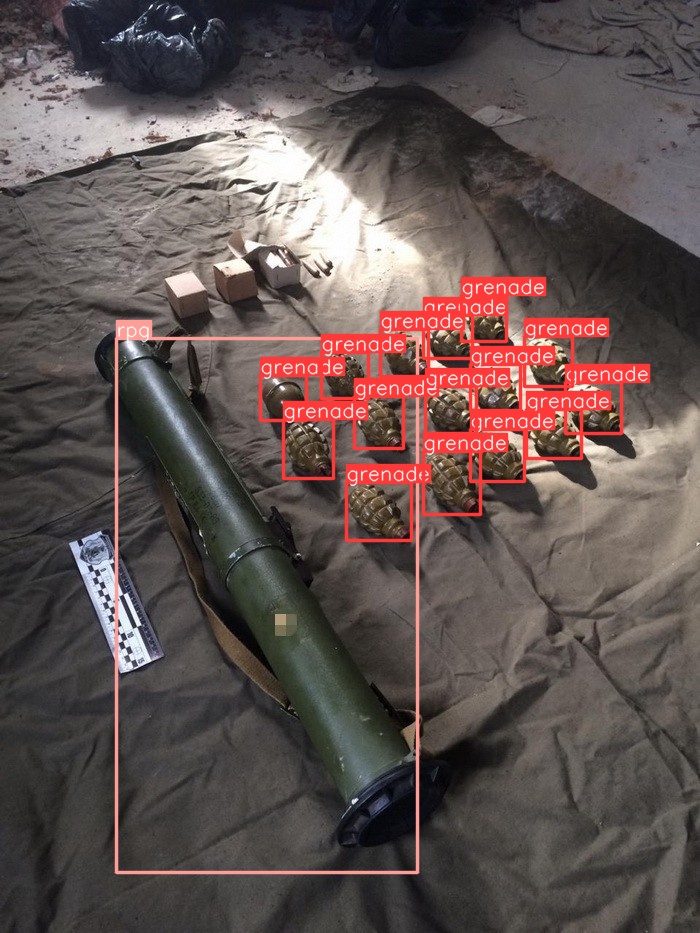} &
\includegraphics[width=.22\textwidth]{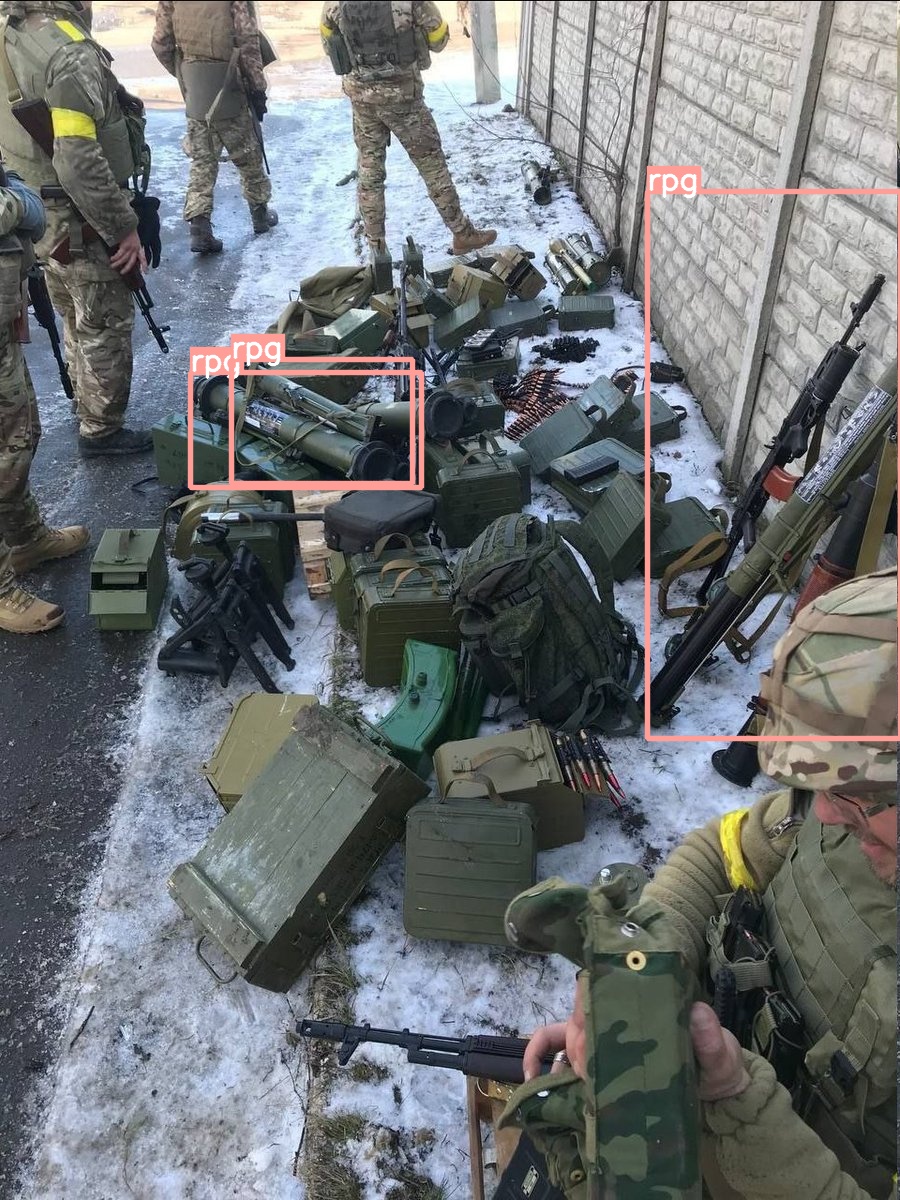} &
\includegraphics[width=.22\textwidth]{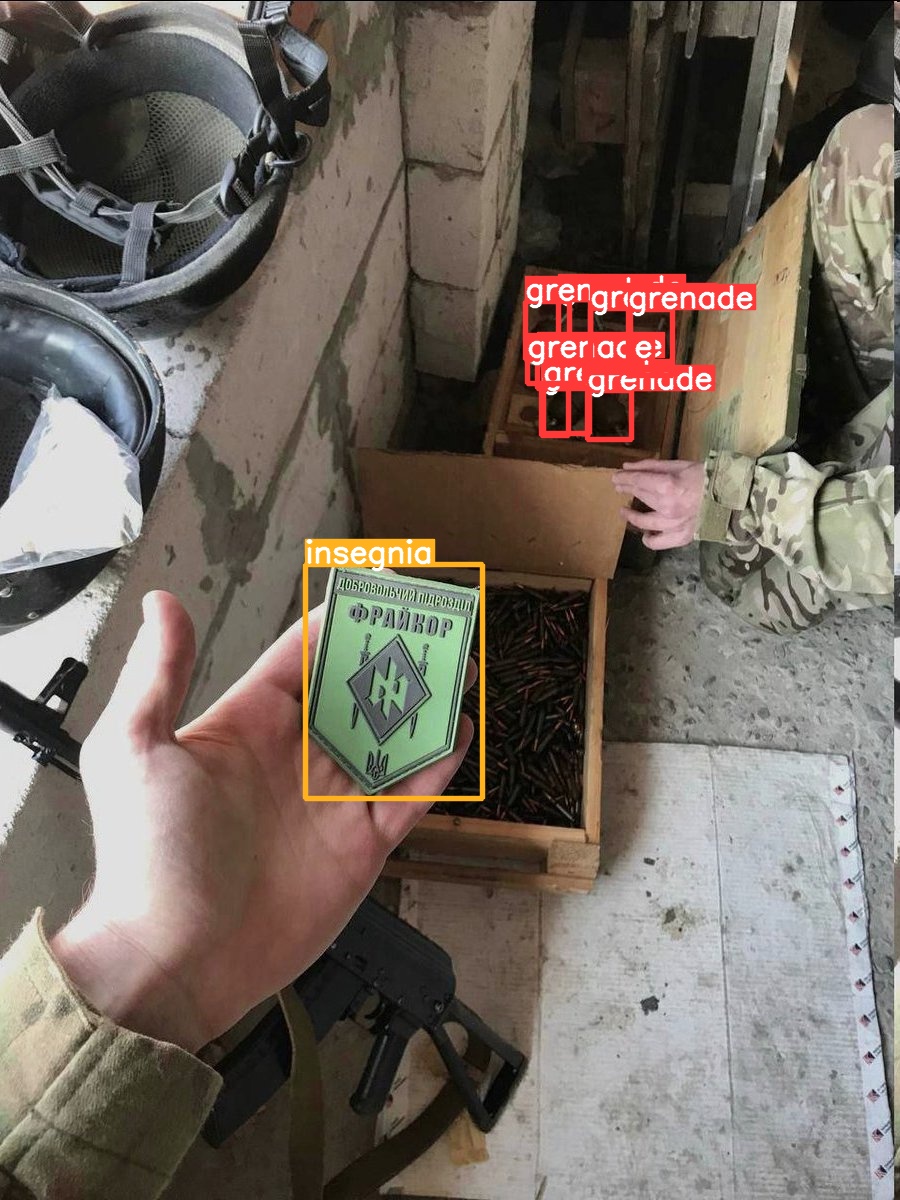} &
\includegraphics[width=.22\textwidth]{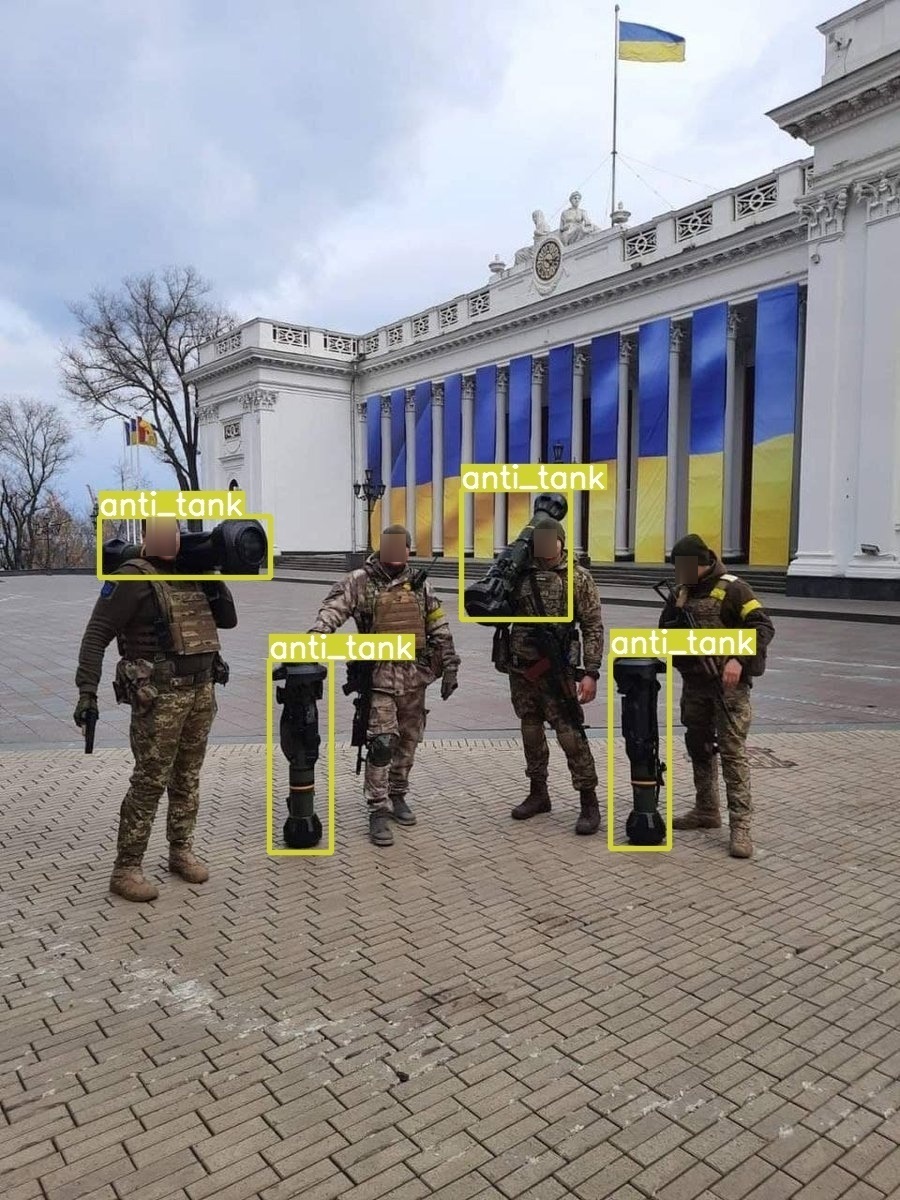} \\

\includegraphics[width=.22\textwidth]{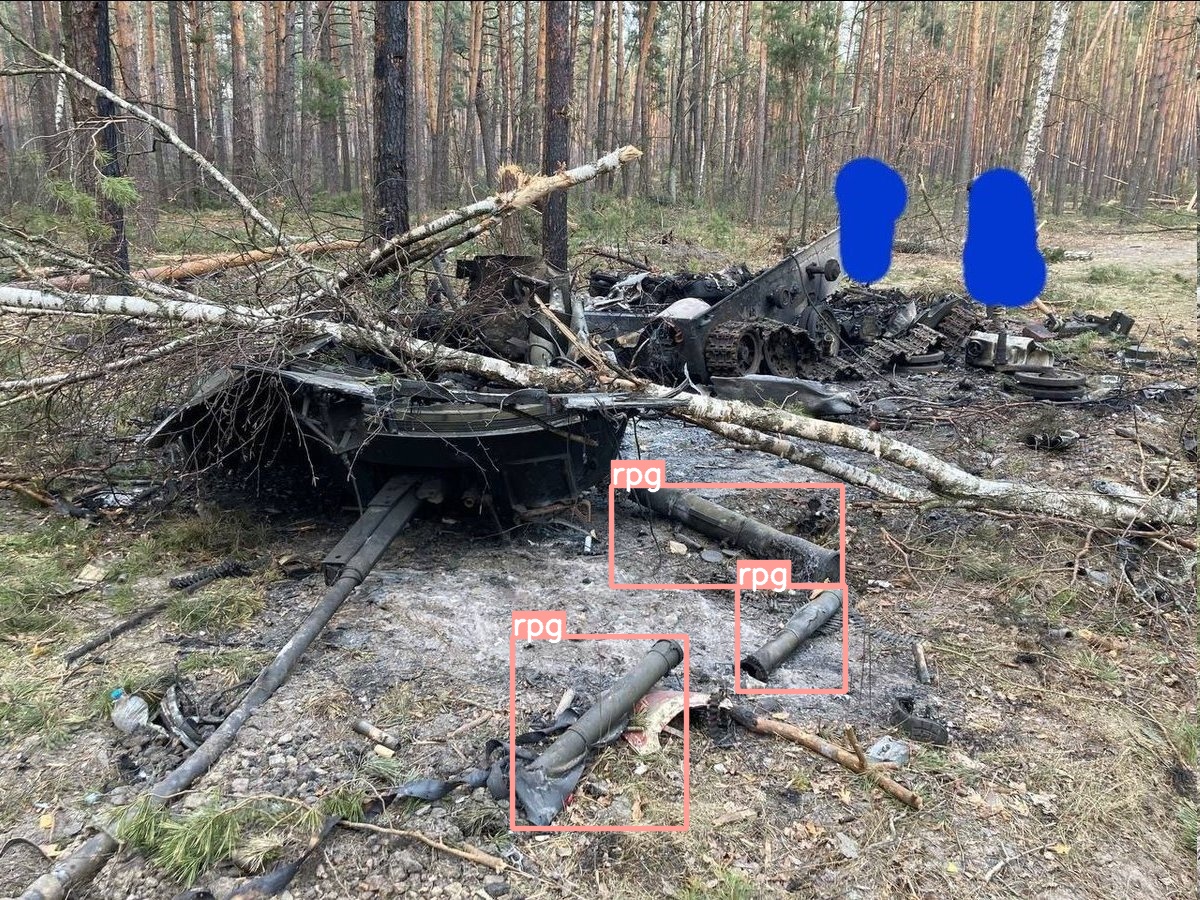} &
\includegraphics[width=.22\textwidth]{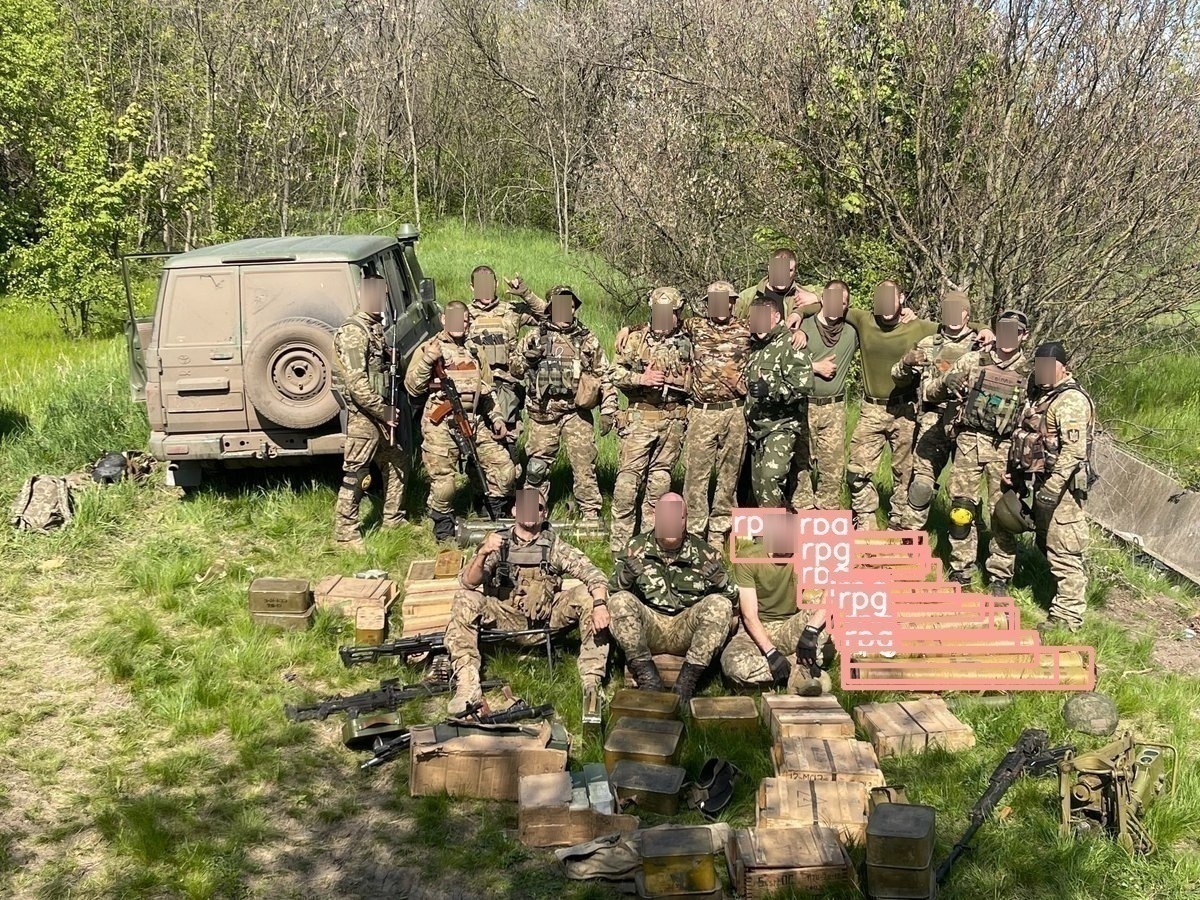} &
\includegraphics[width=.22\textwidth]{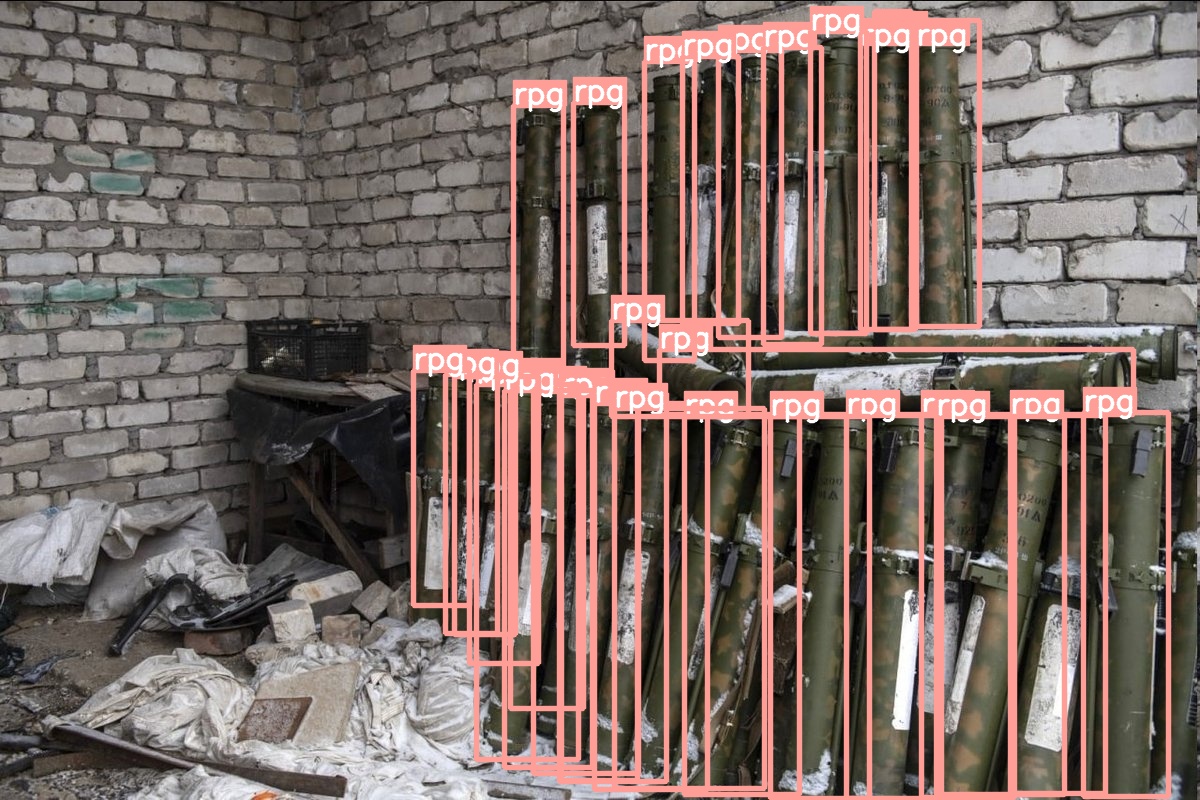} & 
\includegraphics[width=.22\textwidth]{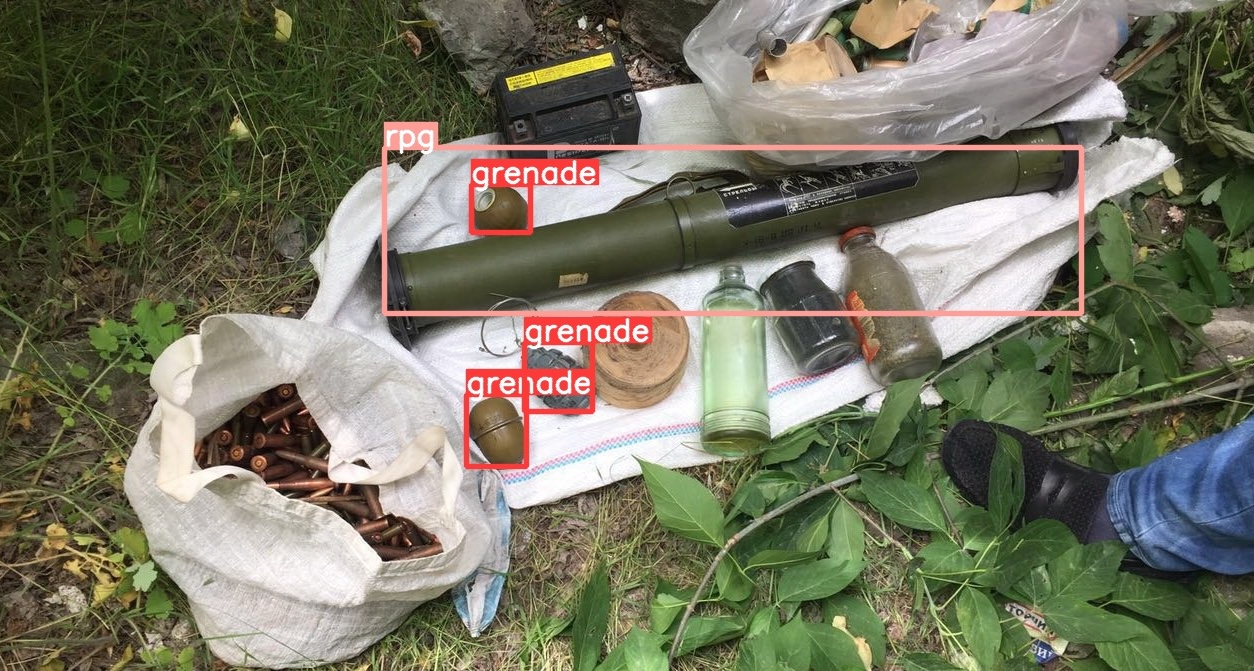}\\

\end{tabular}
\caption{YOLOv11 accurately detects weapons and insignias in challenging situations}
\label{fig:goodresults}
\end{figure*}

Using a testing set of 3,054 images, Table \ref{tab:yolo_perf} shows the performance of YOLOv11 across all image types, sorted by frequency of object.
Results show YOLOv11 tends to perform best for aircraft, tanks, and grenades and worst for civilians and trenches.
If we exclude civilians and trenches, however, as they are very rare in the test dataset, performance increases substantially, with average precision, recall, mAP, and F1 between 0.65-0.68, with lowest performance for insignias.

\subsection{Engagement Across Objects}

\cljb{Audience dynamics on social media platforms can be viewed as feedback loop wherein audiences signal to creators the kinds of content they like via engagement, and creators respond to this signal by creating more of that kind of content \citep{10.1177/1940161220964767.2022}.}
\cljb{We therefore measure the average engagement weapons and war-related content receive relative to the frequency of that class (see Table \ref{tab:engagement}).}
While images of tanks are by far the most frequent---they appear over three times more often than civilian vehicles, the next most frequent class of object---users' engagement shows a different pattern: Images YOLOv11 identifies of warships, while rare, receive substantially more engagement in both retweets and likes, whereas tanks are in the middle engagement-wise.
Qualitative assessment of the log-scale distributions of engagement suggest that the overall distributions of engagement are not be significantly different; that is, their distributions appear log-normal, consistent with other analyses of social media engagement \citep{10.1007/s13278-022-00928-2}. 
Consequently, while raw counts may vary across the various types of weapon systems, audiences may not express a specific preference for individual types of weapons, removing a potential source of confounding when comparing online prevalence to offline events.

\begin{table}[htb]
\caption{Engagement Per Image Type.}
\begin{center}
\begin{tabular}{l r r r}

\textbf{Class}  &  \textbf{n}&  \textbf{Mean Retweets} &  \textbf{Mean Likes} \\ \hline 
Military Warship  & 11 &  555  &  4,282  \\
Civilian Vehicle  & 39 &  498  &  3,772  \\
Anti-Tank  & 67 &  485  &  3,578  \\
Weapon  & 258 &  442  &  3,993  \\
Military Truck  & 406 &  442  &  3,687  \\
Grenade  & 36 &  435  &  3,331  \\
Camouflage Soldier  & 73 &  426  &  3,941  \\
Soldier  & 802 &  393  &  3,870  \\
Military Tank  & 2,960 &  374  &  3,583  \\
Military Artillery  & 123 &  368  &  3,286  \\
Insignia  & 323 &  361  &  3,466  \\
Military Aircraft  & 655 &  359  &  3,370  \\
Civilian  & 9 &  343  &  3,252  \\
Land Mine  & 8 &  334  &  2,960  \\
RPG  & 324 &  324  &  2,818  \\
Military Vehicle  & 277 &  318  &  3,164  \\

\end{tabular}
\end{center}
\label{tab:engagement}
\end{table}

\begin{figure*}[t]
\begin{center}
\includegraphics[width=0.75\textwidth]{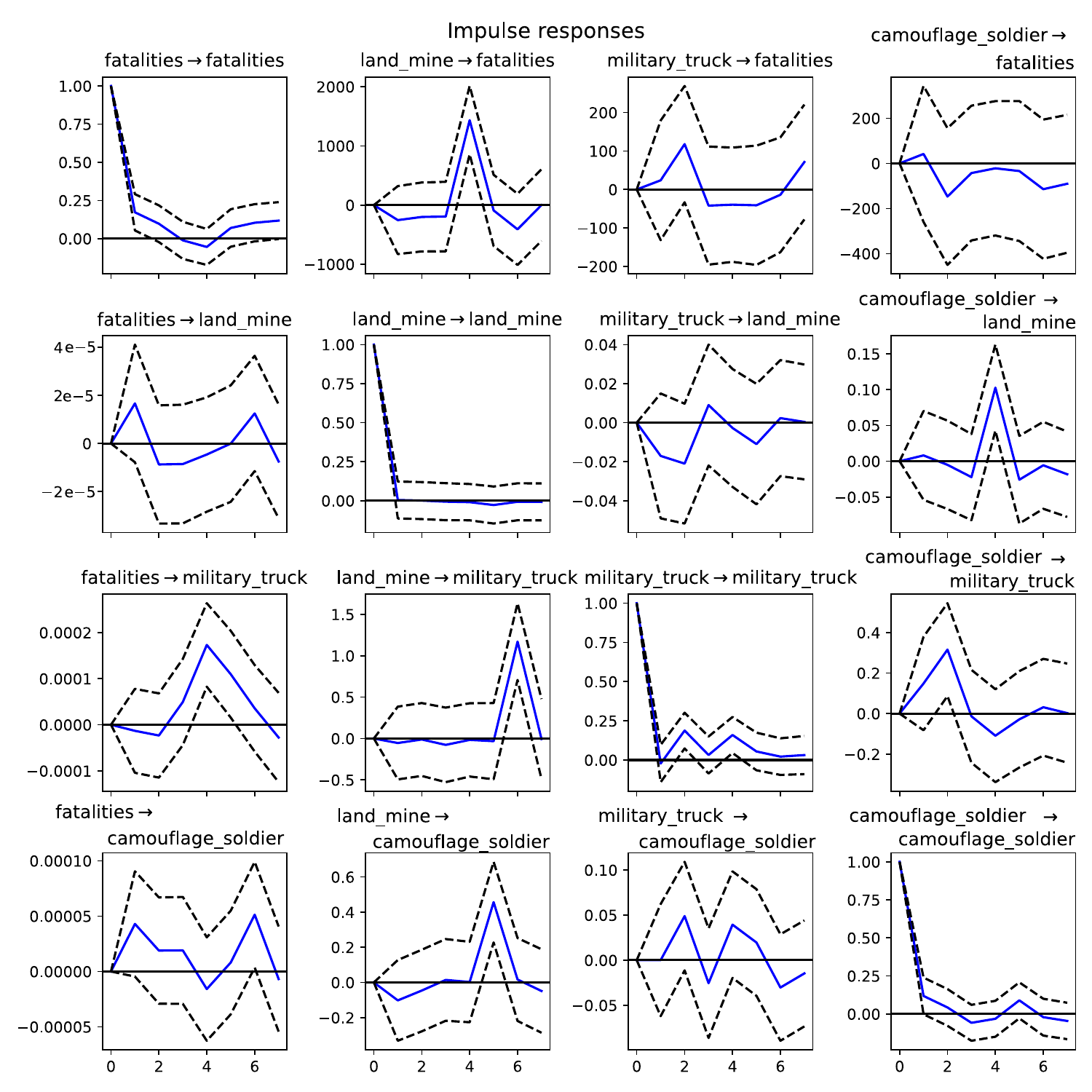}
\caption{These figures show the impulse response functions across our major factors of interest. Each cell is labeled to show the influence of the independent variable on the dependent variable, with the x-axis corresponding to the lag in days between the two factors, while the y-axis shows the expected response (in blue) to a one-unit increase in the independent variable, bounded by a 95\% confidence interval (dashed line). For example, the second cell in the first row shows an increase in images containing landmines corresponds to a significant increase in fatalities four days later.}
\label{fig:normed_irf}
\end{center}
\end{figure*}

\subsection{Assessing Temporal Correlation}

In assessing correlations among time series data from online behavior and offline events, we first check for significant Granger causality between overall post frequency and fatalities/attacks, where we find none.
Moving to correlations between object frequencies and normalized time series data for online behavior, after applying Bonferroni correction $p < (0.05 / 17)$, we find a significant relationship among fatalities and land mines, camouflage soldiers, and military trucks.
These relationships are present only for normalized object time series, with fatalities leading camouflage soldiers and military trucks by one and four days respectively.
For land mines, however, normalized land-mine frequency leads fatalities by four, five, six, and seven days.
We see no significant results for number of attacks and either object frequencies and normalized time series data.

Using the results from our Granger causality analysis, we then fit VAR models to the normalized time series data for these three factors---land mines, camouflage soldiers, and military trucks---to fatality data.
Figure \ref{fig:normed_irf} shows the impulse responses across these factors.
Results from this analysis are largely consistent in that land mines appear to lead fatalities at four days, and fatalities appear to lead military trucks and camouflaged soldiers at four and six days respectively.


\section{Discussion}

As we explore new applications of human-AI teamwork for analyzing conflict dynamics using social media data, we have identified several challenges and avenues for future work.

\subsection{AI Challenges}

This research shows that YOLOv11 can provide encouraging, preliminary success in detecting multiple types of weapons and insignias in real-world environments, even with a relatively small data set. 

\begin{figure}[bht]
\centering
    \begin{subfigure}[t]{0.25\textwidth}
        \centering
        \includegraphics[width=0.95\textwidth]{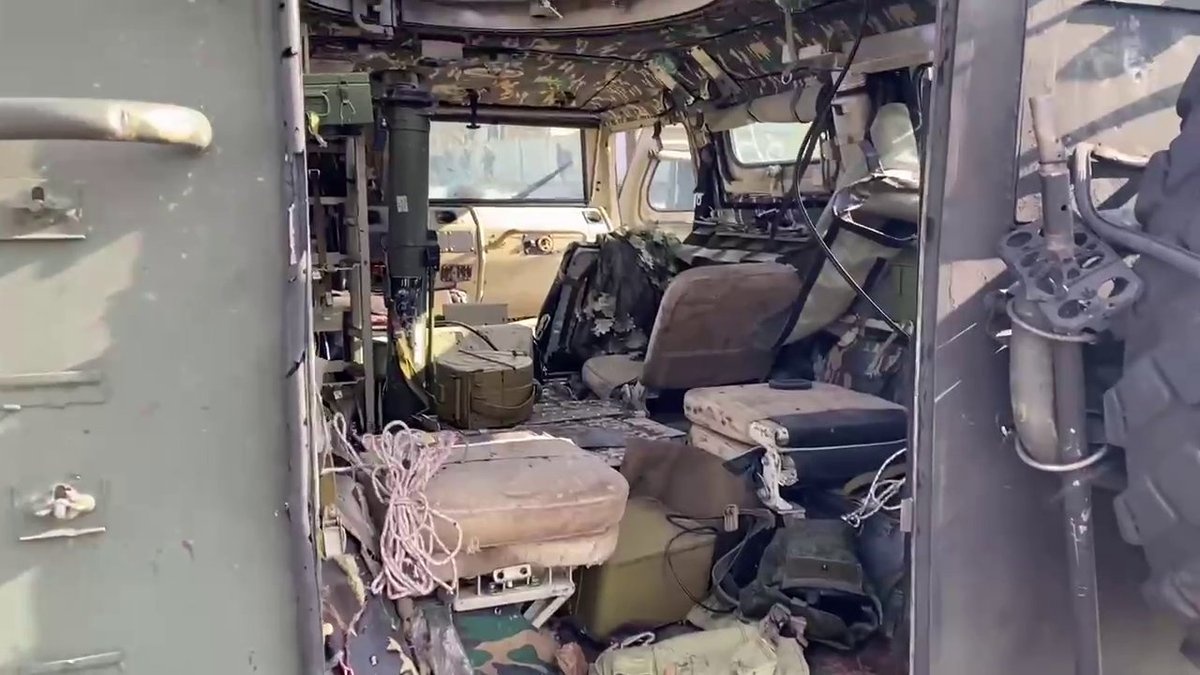}
        \caption{An RPG Inside a Military Vehicle}\label{fig:missdetection_1}
    \end{subfigure}%
    ~ 
    \begin{subfigure}[t]{0.25\textwidth}
        \centering
        \includegraphics[width=0.95\textwidth]{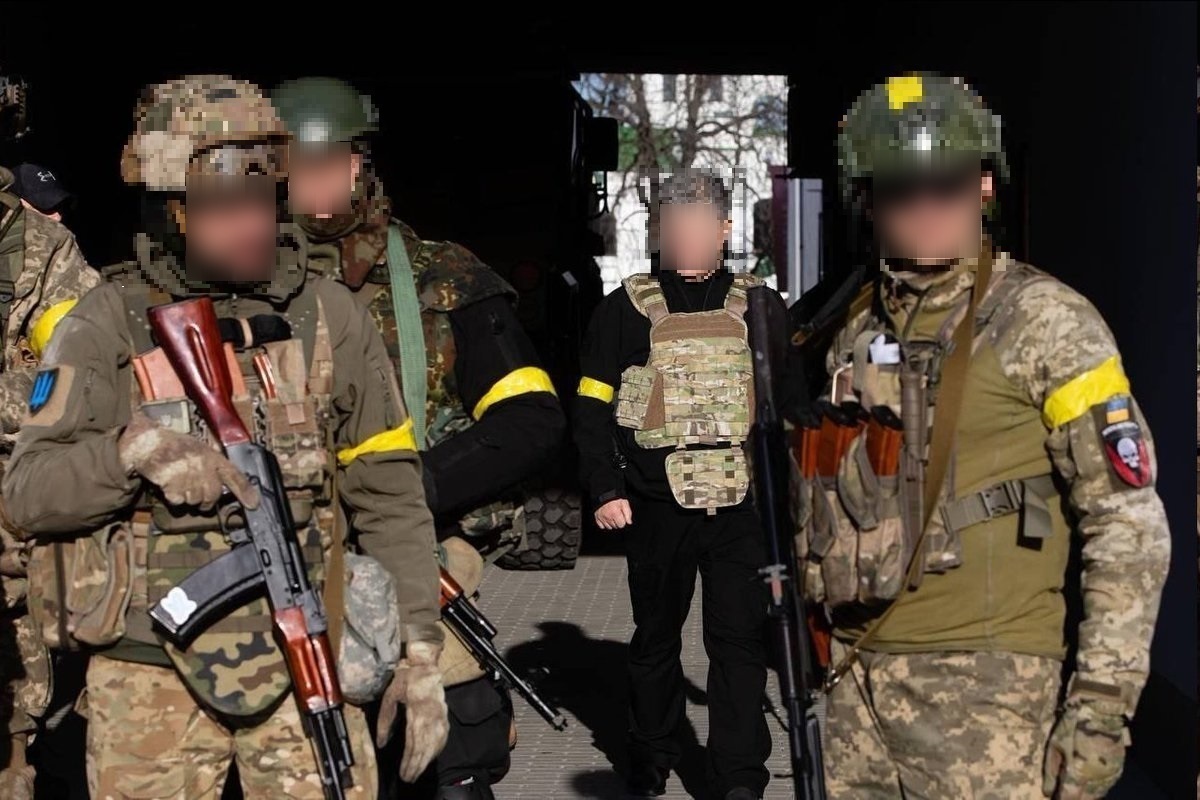}
        \caption{Insignias and Weapons}\label{fig:missdetection_2}
    \end{subfigure}
\caption{Difficult-to-Identify Weapons in Complex Images.}
\label{fig:misdetection}
\end{figure}

\textbf{Improved Accuracy.}
There are still challenges to overcome, as demonstrated by the difficult images. In Figure \ref{fig:missdetection_1}, an RPG is not identified due to background similarity, while Figure \ref{fig:missdetection_2} shows insignias that were missed. 
Future work aims to improve the accuracy and effectiveness of detecting a wider range of weapons and insignias in challenging field settings while improving accuracy.

%

\textbf{Human-AI Teaming.}
The process of analyzing and extracting valuable information with AI will not be entirely automated---for example, in this research, we had to collect an image data set and arms experts helped us determine what we should look for in our collected images. However, the more accurate our tools are, the easier it will be for human users to validate and trust the data output from the system. Additionally, it will be up to human subject matter experts to identify an ever-changing list of what types of objects should be automatically identified and how the identified data will be used.

\textbf{Multi-Modal Analysis.}
In addition to static images, many conflict groups post videos of themselves with their weapons \citep{warSM}.
Analysis of these videos may identify both objects of interest and the organizational size and geographic location of various armed groups. 
Language can be automatically analyzed, whether from audio in a video, from printed material in an image, or in data from associated posts.
Meta-data from posts may also be useful for analysis, including time and date, geotagging information, and network-related information (e.g., re-posting or responding).

\textbf{Real-time vs.\ Post Hoc Analysis.}
AI can help us gain a better spatiotemporal understanding of how conflicts evolve over time. Currently, analysts can perform AI-assisted post hoc analysis, helping us better understand the evolution and social networks of armed actors by tracking weapon movement between groups and across a conflict zone. Eventually, such analysis could be conducted in real time, allowing analysts to understand changing dynamics and make timely suggestions.

\subsection{Policy Impacts}

This work suggests new potential for human-AI teamwork to analyze social media data, providing novel and relevant analysis for active conflict settings. 
Systems that can help quickly identify and relay critical analysis of emerging events and threats in conflict zones have substantial value in development and humanitarian aid work. 
In particular, owing to the intensity of violence, real-time information from the ground is often difficult to collect due to the risk faced by fieldworkers, and social media channels provide important, if imperfect, solutions to data shortages. 
Future analysis can help researchers parse the information and label different locations, people, dates, and, most importantly, the relationship between entities. For example, when a particular location is relevant to multiple posts, these data can prove invaluable in identifying military flashpoints. This information has practical implications for humanitarians as it could provide early warning of operational significance (e.g., populations displaced by fighting or threats to humanitarian staff in the field).
Furthermore, in civil war contexts, how weapons diffuse through networks of armed groups provides unique information about the relationship between those groups. This information is crucial for identifying alliances between armed groups and who controls specific transportation routes or cities. Understanding the landscape of armed groups is essential for humanitarian agencies to identify with whom to negotiate to gain access to communities in need.  However, the benefits of this analysis must be weighed against its potential risks, as it has dual-use implications. Malicious actors could employ similar methods to track weapon diffusion networks or anticipate future military operations, raising ethical concerns about the dissemination of this and related methodologies.

As the potential for conflict increases, either from political instability, climate change, or great-power competition, methods and tools that can take advantage of the increasing documentation of these events are likely to prove crucial.
The Ukraine war, the conflict in Gaza between Israel and Hamas \citep{Dostri:2024aa}, and the continued conflict in Syria \citep{bachmann2021syria} are all examples of conflicts that have had substantial coverage in online spaces, and such conflicts are not likely to diminish in the near future.

\section{Conclusion}
This paper has introduced novel uses of computer vision to analyze social media posts for open-source intelligence and conflict studies. 
As this transdisciplinary work progresses, opportunity exists to build new tools that support conflict experts, humanitarian workers, and other stakeholders in understanding conflict and its evolution. 
Humanitarian organizations are calling for more tools to help them harness the potential of social media data in conflict zones in an ethical way that protects the privacy and dignity of those impacted by conflict. 
This includes combating attempts by malicious actors to target them using campaigns of disinformation and direct threats of violence.  
There is also potential to track the targeting and impact of military attacks on healthcare facilities, both from the perspective of identifying communities in need of medical aid and documenting the extent of violations of humanitarian law.The ability to track this information holds the potential to be harnessed by malicious actors as well, raising additional ethical concerns for researchers and granting access to their methods. 
Such tools need to respect the privacy and dignity of those impacted by conflict, but success will bring many new abilities, including identifying communities in need of medical and humanitarian aid, documenting violations of humanitarian law, and fighting disinformation.

\begin{acks}

Research was sponsored by the Army Research Office and was accomplished under Grant Number W911NF-23-1-0013. 
The views and conclusions contained in this document are those of the authors and should not be interpreted as representing the official policies, either expressed or implied, of the Army Research Office or the U.S. Government. 
The U.S. Government is authorized to reproduce and distribute reprints for Government purposes notwithstanding any copyright notation herein.
Part of this work has taken place in the Intelligent Robot Learning Lab at the University of Alberta, which is supported in part by research grants from Alberta Innovates; Alberta Machine Intelligence Institute (Amii); a Canada CIFAR AI Chair, Amii; Digital Research Alliance of Canada; Mitacs; and the National Science and Engineering Research Council.
\end{acks}

\bibliographystyle{ACM-Reference-Format}
\bibliography{references}

\end{document}